\documentclass[floatfix,amssymb,preprint,aps, superscriptaddress,showpacs]{revtex4}
\usepackage{amsmath}
\usepackage{amsfonts}
\usepackage{amssymb}
\usepackage{graphicx}
\usepackage{units}
\usepackage{color}
\usepackage{xspace}
\usepackage[normalem]{ulem}
\usepackage{natbib}
\usepackage{textcomp}
\usepackage{gensymb}
\usepackage{romannum}

\newcommand{\fig}[1]{Fig.~\ref{#1}}

\DeclareUnicodeCharacter{2009}{\,}
\DeclareUnicodeCharacter{2212}{-}

\begin{document}

\title{Electronic access to glass transition in supercooled ionic liquids using ambipolar transistor}

\author{Tanima Kundu}
\affiliation{School of Physical Sciences, Indian Association for the Cultivation of Science, 2A $\&$ B
Raja S. C. Mullick Road, Jadavpur, Kolkata - 700032, India
}

\author{Rahul Paramanik}
\affiliation{School of Physical Sciences, Indian Association for the Cultivation of Science, 2A $\&$ B
Raja S. C. Mullick Road, Jadavpur, Kolkata - 700032, India
}

\author{Aishee Saha}
\affiliation{School of Physical Sciences, Indian Association for the Cultivation of Science, 2A $\&$ B
Raja S. C. Mullick Road, Jadavpur, Kolkata - 700032, India
}

\author{Shibnath Mandal}
\affiliation{School of Physical Sciences, Indian Association for the Cultivation of Science, 2A $\&$ B
Raja S. C. Mullick Road, Jadavpur, Kolkata - 700032, India
}

\author{Antarik Sarkar}
\affiliation{School of Physical Sciences, Indian Association for the Cultivation of Science, 2A $\&$ B
Raja S. C. Mullick Road, Jadavpur, Kolkata - 700032, India
}

\author{Bipul Karmakar}
\affiliation{School of Physical Sciences, Indian Association for the Cultivation of Science, 2A $\&$ B
Raja S. C. Mullick Road, Jadavpur, Kolkata - 700032, India
}

\author{Subhadeep Datta}
\email{sspsdd@iacs.res.in}
\affiliation{School of Physical Sciences, Indian Association for the Cultivation of Science, 2A $\&$ B
Raja S. C. Mullick Road, Jadavpur, Kolkata - 700032, India
}

\begin{abstract}

\textbf{Relaxation dynamics of supercooled liquids approaching glassy arrest remain a central challenge in integrated electronic architectures, where conventional rheometry becomes incompatible. Here, we demonstrate that an ambipolar PdSe$_2$ field-effect transistor functions as an electrical probe capable of resolving ion-specific relaxation dynamics in fragile ionic glass formers and semiquantitatively inferring rheological parameters within an operating device environment. Temperature evolution of the transfer curve hysteresis and time-resolved current transients under ionic-gate pulse reveal a non-Arrhenius fragile slowdown. We track the continuous reduction of dynamically equilibrated liquid regions approaching the glass transition through an electrically accessible quantity $p_\text{eq}(T)$, quantifying the fraction of the mobile ions able to relax within the experimental timescale. Upon cooling, $p_\text{eq}$ collapses sharply as mobile regions fragment into percolating fractal clusters, consistent with a reduction of configurational entropy predicted for fragile glass formers. This approach enables temperature-dependent scaling of viscosity and extraction of characteristic temperatures marking the ergodic-to-nonergodic crossover, within a solid-state device architecture where conventional rheological characterization is inapplicable. Further, polymer confinement of the ionic liquid shifts these characteristic temperatures upward, demonstrating the sensitivity of this method to structural constraints imposed by the polymer matrix.}

\end{abstract}

\maketitle

\section{Introduction}

At the most elementary level, the viscosity of a liquid is quantified by measuring the resistance of a fluid to flow under an applied stress, as in classical capillary or falling-sphere experiments. Extending such rheological characterization to supercooled liquids which are metastable, structurally disordered fluids below their equilibrium melting temperature, remains challenging, particularly for their temperature-dependent viscosity, relaxation spectra, and approach to glass transition, yet such knowledge is essential in soft condensed matter, electrochemistry, nanoscale electronics, and glass physics \cite{Angell1995Science,Ediger1996JPC,Shakeel2019ACS,Mauro2009pnas}. Conventional rheometric techniques, e.g., rotational shear, oscillatory methods, or microfluidic viscometry, demand at least milliliter volumes, and mechanical stability \cite{Shakeel2019ACS,Larson1999oxford} that are incompatible with strong confinement or solid-state device architectures \cite{Zuo2019pccp,Han2022Langmuir}. A method that can probe the microscopic dynamics of supercooled liquids in tiny volumes and without any mechanical actuation would overcome a long-standing experimental gap. 

Ionic liquids constitute an intriguing class of supercooled liquids, combining negligible vapor pressure ($<10^{-10}$~Pa at 300~K), high ionic conductivity ($10^{-3}$--$10^{-1}$~S\,cm$^{-1}$), and a wide electrochemical stability window ($\delta V \approx 4-6$~V), while retaining pronounced dynamical heterogeneity near room temperature unlike conventional molecular glass formers \cite{Hayes2015ChemRev,Paulechka2003J.Chem.Eng.Data,Pereira2024ionics,Pachernegg2024CED}. When employed as gate dielectrics, they form nanometer-thick electric double layers capable of inducing surface charge densities exceeding 10$^{14}$ cm$^{-2}$ \cite{Ye2011pnas,Fujimoto2013pccp,Cheng2022nanolett}. This exceptional electrostatic efficiency enables phase modulation, insulator-to-metal transition, and even superconducting transition in van der Waals materials \cite{Cheng2022nanolett,Saito2015ACSNano,Wang2023natcom,Jo2015nanolett,Chen2018advmat}. Beyond their technological utility, ionic liquids represent a unique state of matter: structurally disordered, dynamically heterogeneous, and capable of undergoing glasslike freezing \cite{Angell1995Science,Ediger1996JPC,Grzybowski2015scirep,Lubchenko2015AdvPhys}. However, their extreme sensitivity to temperature gradients and interfacial interactions renders conventional rheology ineffective, particularly when integrated into thin films, nanofluidic geometries, or polymer network \cite{Shakeel2019ACS,Liu2024Macromol}. 

Here, we demonstrate that an ambipolar ionic-liquid–gated PdSe$_2$ transistor functions as an electrical probe capable of transducing ion dynamics directly into electronic transport. PdSe$_2$ exhibits intrinsic ambipolar conduction, enabling access to both electron and hole transport [a maximum mobility ratio obtained for multilayered PdSe$_2$, $(\mu_e/\mu_h)_\text{max}\approx6$] \cite{Oyedele2017JACS}. The channel conductance follows the instantaneous electrostatic potential as $G(t)=G_0+\alpha\left[V_g-\phi_{\mathrm{EDL}}(t)\right]$, where $\phi_{\mathrm{EDL}}$ relaxes via overdamped ionic dynamics described by the Poisson-Nernst-Planck framework~\cite{Das2008NatNano,Palaia2025PRE}. Finite-rate gate sweeps induce hysteresis due to delayed ionic relaxation, linking the hysteresis width $\Delta V_H$ to the ionic mobility $\mu_{\mathrm{ion}}$. The temperature dependence of $G(t)$ and $\Delta V_H$ therefore yields the ionic relaxation time $\tau_{\mathrm{ion}}$, from which viscosity can be semiquantitatively inferred, enabling electrical characterization of ionic dynamics in nanoelectronic volumes.

We apply this approach to the prototypical ionic liquid EMIM-TFSI \cite{Ishisone2022pccp}, studied both as a pure liquid and as a polymer-confined ionogel with PVDF-HFP \cite{Yuan2015Springer, Yang2017ACSAMI, Lee2012AdvMat}. EMIM-TFSI with its well-established physiochemical properties is chosen as a model system to benchmark our electrical measurements directly against the established literature. The ambipolar character of PdSe$_2$ allows polarity-selective gating, enabling the relaxation kinetics of EMIM$^+$ and TFSI$^-$ to be resolved independently. Temperature-dependent hysteresis broadening in the transfer characteristics, from nearly zero at room temperature to $\Delta V_\text{H}^{200\,\mathrm{K}} \approx 4.5\,\mathrm{V}$, together with current transient measurements under ionic-gate voltage steps ($0 \rightarrow +3\,\mathrm{V}$ and $0 \rightarrow -2.5\,\mathrm{V}$), reveal a non-Arrhenius slowdown consistent with fragile glass-forming behavior \cite{Angell1995Science, Böhmer1993JCP,Coslovich2007JCP}. The extracted relaxation time grows from $\sim 1\,\mathrm{s}$ near room temperature to $>100\,\mathrm{s}$ approaching $200\,\mathrm{K}$. Concurrently, the fraction of regions that can fully equilibrate within the experimental timescale at high temperature decreases continuously on cooling ($\sim 50\%$ around 213 K for pure EMIM-TFSI), demonstrating the onset of ergodicity loss \cite{Comez2005PRL,Costa2022NatPhy}. The reduction of this mobile fraction implies a progressive loss of connectivity among relaxation pathways. In percolative descriptions of glassy dynamics, such sparse connectivity leads to ramified clusters with reduced effective dimensionality, consistent with fractal-like transport pathways~\cite{Wool2008JPS, Ma2008NatMat}. Within the Adam-Gibbs framework, the dramatic increase in relaxation time reflects a depletion of configurational entropy defined by the number of distinct structural arrangements available to the liquid~\cite{Adam1965JCP}. The measured electrical relaxation infers Vogel-Fulcher-Tammann (VFT) scaling of viscosity and captures the progressive reduction of configurational degrees of freedom. Polymer-confined ionogel shifts the characteristic temperatures upward by $\sim 15$--$30\,\mathrm{K}$ and substantially increases the viscosity relative to the neat liquid, consistent with reduced free volume and confinement-enhanced suppression of cooperative mobility. Altogether, this study introduces ambipolar PdSe$_2$ transistor as a device-integrated platform for electrically probing the rheology and glass physics of supercooled ionic media, where conventional rheometry cannot operate.

\section{Methods}

{\it Materials.} Standard self-flux growth technique was adopted to grow single crystals of PdSe$_2$ followed by structural solution and stoichiometric characterizations as described in ref. \cite{Kundu2023chemmat}. 1-ethyl-3 methylimidazolium bis(trifluoromethylsulfonyl)imide, [EMIM][TFSI] (high purity) ionic liquid and poly(vinylidene fluoride-co-hexafluoropropylene), P(VDF-HFP) were purchased from Sigma Aldrich and stored in a nitrogen-filled glovebox.

{\it Ionogel preparation.} The free-standing ionogel layer was obtained by spin-coating the precursor solution on a cover slip. The precursor solution was prepared by sequentially dissolving PVDF-HFP and EMIM-TFSI in acetone followed by stirring the solution at 50\degree C for 2 h in a nitrogen-filled glovebox. The weight ratio among the polymer, IL, and the solvent was kept to 1:4:7. The spin-coated ionogel was kept in a vacuum oven at 70\degree C for 3 h to remove the residual solvent.
  
{\it Device fabrication and transport measurements.} Multilayered PdSe$_2$ flakes were mechanically exfoliated onto 285 nm SiO$_2$ coated $p$-Si substrates. Optical lithography was used to pattern source-drain contact electrodes along with local gate electrodes followed by Ti(8 nm)/Au (60 nm) metallization using e-beam evaporation. Performance of the PdSe$_2$ field-effect transistor (FET) was examined by employing ionic liquid gating as a superior alternative of SiO$_2$ gate. A droplet of ionic liquid (IL) was applied using a micropipette covering the prepatterned three-terminal FET device of the PdSe$_2$ channel inside an argon filled glove box. To minimize the air exposure, the device was transferred to the measurement chamber immediately after dropcasting the IL. On the other hand, the spin-coated copolymer ionogel was cut with a razor blade and then laminated onto the semiconductor device using a tweezer. A gold wire served as a top gate contact for the ionogel film. Electrical transport measurements were conducted inside a closed cycle cryostat (ARS-4HW) under high vacuum condition (10$^{-6}$ mbar) using Keithley 2601B, 2450, and 2636B source measure units (SMU).

\section{Results and Discussions}
\begin{figure*}[t]
\centerline{\includegraphics[width=\textwidth, clip]{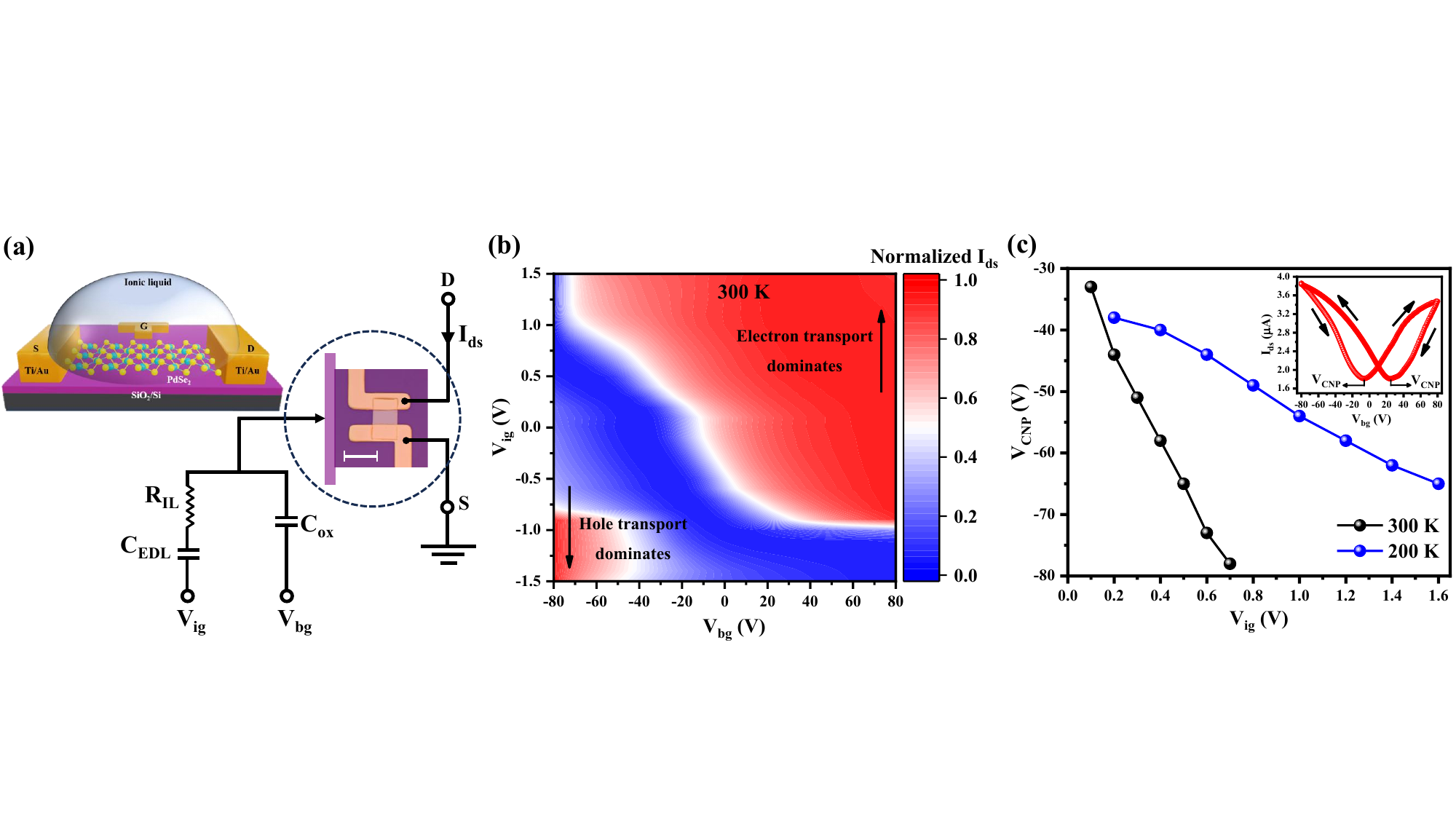}}
\caption{(a) Device architecture of an ionic-liquid gated PdSe$_2$ field-effect transistor with the corresponding equivalent circuit, in which the electric double-layer capacitance of ionic liquid $C_{\text{EDL}}$ and backgate oxide capacitance $C_{\text{ox}}$ act in parallel. $R_{\text{IL}}$ denotes the resistance of ionic liquid. The optical micrograph of the two-terminal PdSe$_2$ channel is embedded within the circuit representation (scale bar $\sim$ 20 $\mu$m). (b) Color map of source–drain current ($I_{\text{ds}}$) as a function of backgate voltage ($V_{\text{bg}}$) at different ionic gate bias ($V_{\text{ig}}$). The arrows indicate the gate-dependent evolution of ambipolar transport, illustrating the tunability of electron and hole conduction upon reversal of the ionic gate polarity. (c) Backgated (forward sweep) charge neutrality point ($V_\text{CNP}$) plotted as a function of ionic gate voltage ($V_\text{ig}$) at room temperature and 200 K, manifesting the freezing of ionic motion near the glass transition regime. Inset shows a representative ambipolar transfer curve indicating two charge neutrality points along forward and backward sweeps. 
\label{schematic}}
\end{figure*}

Multilayer PdSe$_2$ flakes exfoliated from the bulk single crystal were transferred onto SiO$_2$ (285 nm)/$p$-Si substrates, followed by fabrication of Ti/Au contacts \cite{Kundu2023chemmat}. Note that details of crystal growth, structural characterization, and charge-transport properties of PdSe$_2$ from the same batch of samples have been reported elsewhere \cite{Kundu2023chemmat, Kundu2025PRB}. The devices were subsequently gated using SiO$_2$/Si back-gate and EMIM-TFSI ionic liquid [\fig{schematic}(a)]. The ionic bias forms a nanometer-thick electric double layer (EDL) with $C_{\text{EDL}}\sim$ $\mu$F cm$^{-2}$ \cite{Fujimoto2013pccp}, exceeding SiO$_2$ capacitance ($C_{\text{ox}}$). An equivalent circuit model shown in \fig{schematic}(a)) depicts the combined action of the EDL capacitance ($C_{\text{EDL}}$) and the backgate oxide capacitance ($C_{\text{ox}}$). The resulting electrostatic tunability is manifested in the color map in \fig{schematic}(b) generated by the backgated transfer curves ($I_{\text{ds}}$ vs $V_{\text{bg}}$) measured under different ionic-gate biases ($V_{\text{ig}}$), demonstrating that both electron and hole conduction can be continuously tuned by the polarity and magnitude of the ionic-gate voltage. The corresponding stack plot shown in Fig. S1 of the Supplemental Material \cite{supple} clearly portrays the evolution of backgated ambipolar transport towards hole-dominated (\emph{p} type) transport under negative ionic-gate bias and electron-dominated (\emph{n} type) transport under positive ionic-gate bias at room temperature, which is basically caused by the accumulation of positive (negative) ions at the channel interface when $V_{\text{ig}}$ $<$ 0 ($V_{\text{ig}}$ $>$ 0). The backgated charge neutrality point (forward sweep), $V_\text{CNP}$, shows strong tunability with the ionic gate, $V_{\text{ig}}$, at room temperature, but this tunability is substantially suppressed near 200~K due to ionic freezing [\fig{schematic}(c)].

\begin{figure}[t]
\centerline{\includegraphics[width=\textwidth, clip]{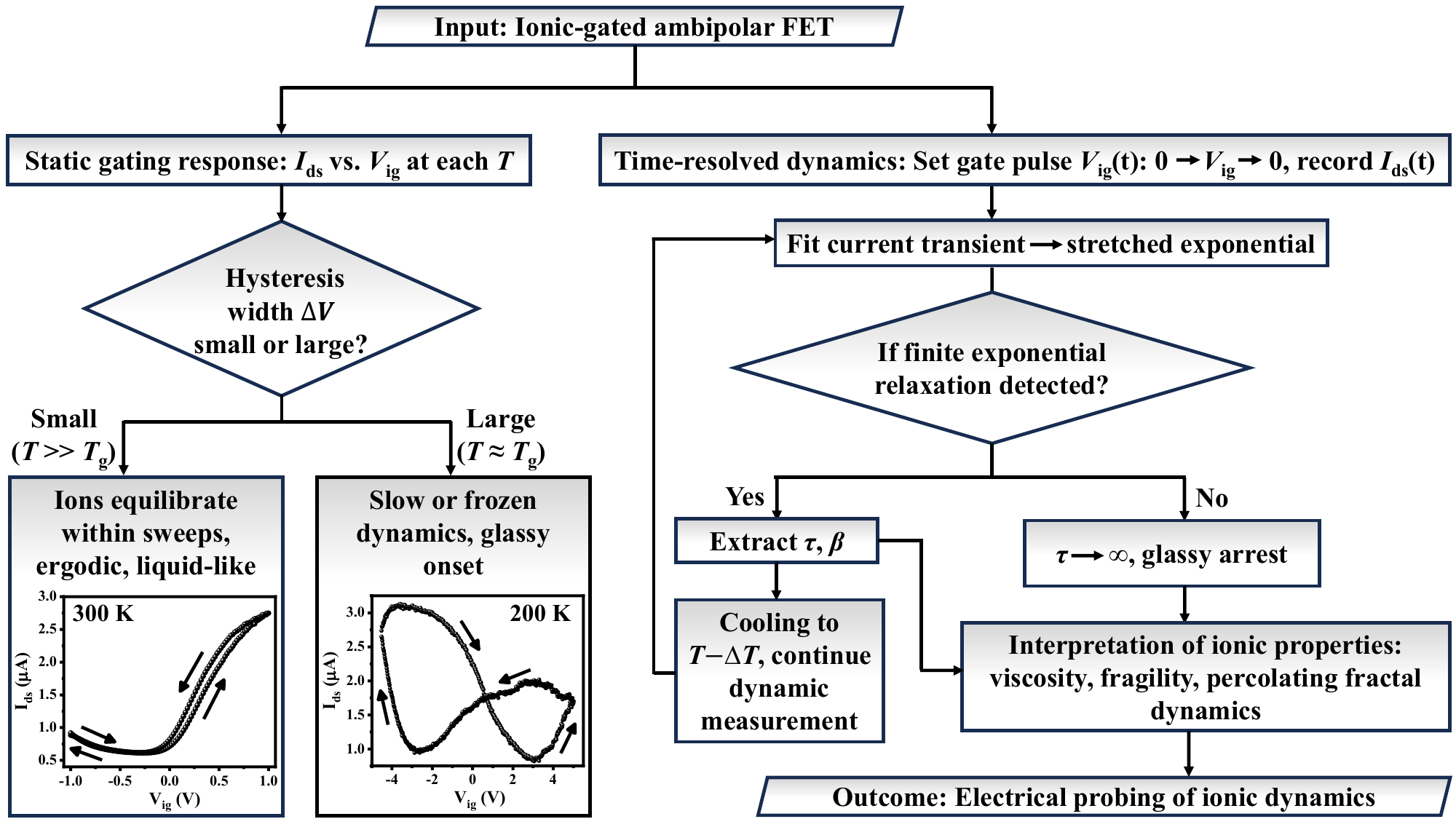}}
\caption{Schematic flowchart outlining the combined static and dynamic measurement protocol used to probe the ionic relaxation of EMIM-TFSI electrically.
\label{flowchart}}
\end{figure}

To elucidate the ionic relaxation mechanisms that govern electrostatic doping, we established a combined static and dynamic measurement protocol. \fig{flowchart} describes the workflow for extracting ionic dynamics in an ionic-liquid-gated ambipolar FET. Static sweeps ($I_{\text{ds}}$ vs $V_{\text{ig}}$) at different temperatures quantify hysteresis $\Delta V_\text{H}$, reflecting ionic relaxation: narrow loops at 300~K and strongly broadened near 200~K due to slow or frozen ions (Fig.~S2 \cite{supple}; see the Supplemental Material for details). Complementary dynamic measurements apply gate-voltage pulses and fit $I_{\text{ds}}(t)$ to a stretched exponential to extract time constant, $\tau$, and stretch exponent, $\beta$. Divergence of $\tau$ signals glassy arrest, enabling interpretation of viscosity, fragility, and fractal ionic transport.

Drain-current transients [$I_{\text{ds}}(t)$] were measured following abrupt ionic-gate voltage steps, predominantly reflecting ionic relaxation \cite{Li2025AdvMatTech,Zhao2024NanoMicroLett}. In a minimal electrical description, the transport-limited charging of the EDL is characterized by the circuit time constant $\tau_\text{e}=R_{\text{IL}}C_{\text{EDL}}$, where $R_{\text{IL}}$ denotes bulk ionic resistance. Beyond this electrical picture, the ionic liquid exhibits viscoelastic relaxation associated with viscosity ($\eta$), a measure of the liquid's response to a suddenly imposed shear stress ($\zeta$) described by the Maxwell spring-dashpot model with $\tau_\text{r}=\eta/\zeta$. Consequently, the experimentally observed drain-current transient does not reflect a single universal time constant, but rather the interplay between transport-limited EDL charging and viscoelastic relaxation in the ionic liquid.

\begin{figure*}[t]
\centerline{\includegraphics[width=\textwidth, clip]{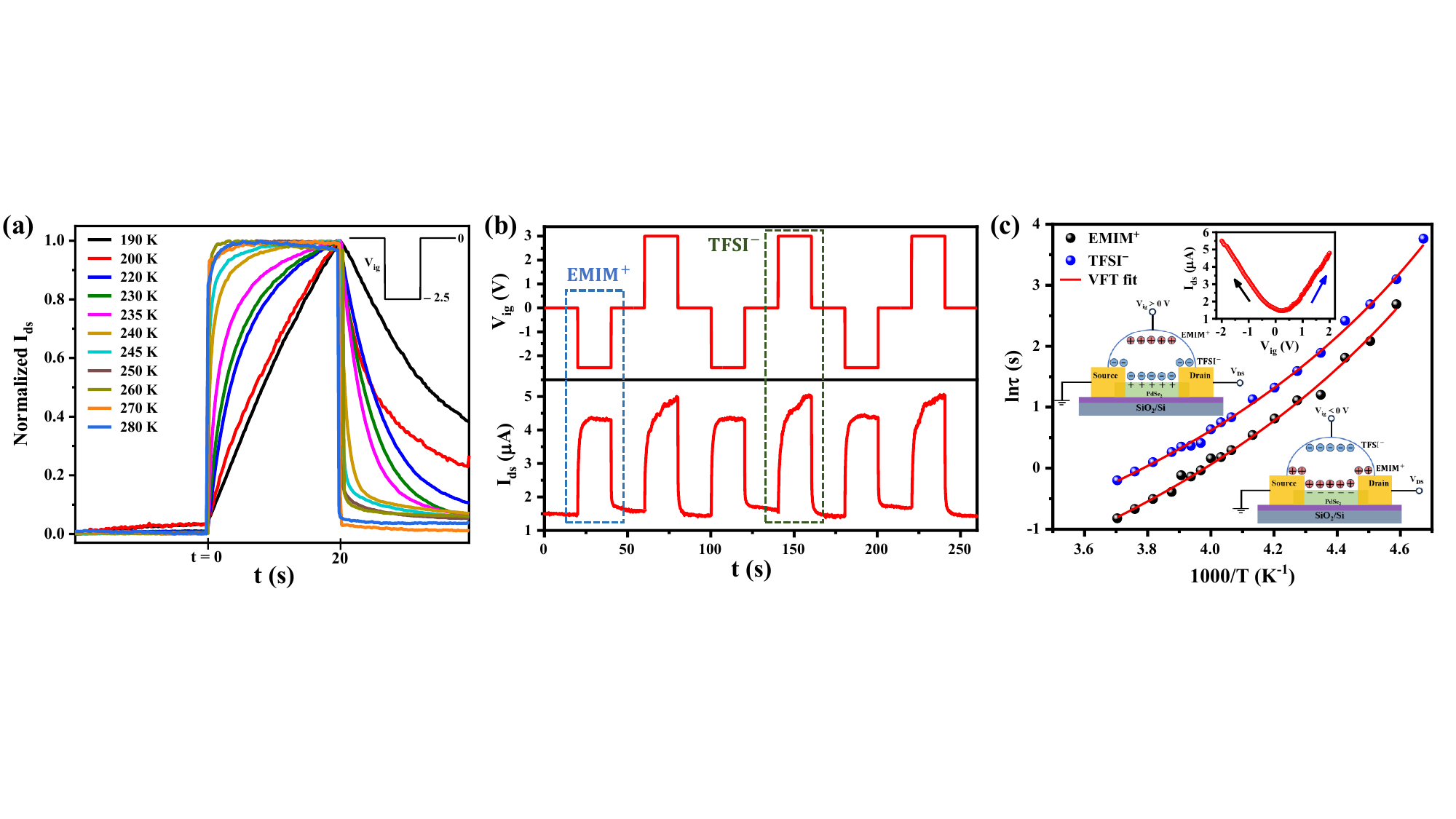}}
\caption{(a) Temperature-dependent drain-current transients measured following a negative ionic-gate pulse (0 $\rightarrow$ –2.5 V) reveal a progressive slowdown of the ionic response upon cooling. (b) Alternating negative and positive gate-voltage steps probe the relaxation of EMIM$^+$ and TFSI$^-$, respectively. The bulkier TFSI$^-$ exhibits a visibly slower relaxation, reflecting reduced mobility. (c) Relaxation times $\tau$ extracted from current transients follow VFT scaling as graphed by $\ln\tau$ vs 1000/$T$, with $\tau_{\text{TFSI}^-}$ consistently exceeding $\tau_{\text{EMIM}^+}$. Insets show an representative ambipolar transfer curve and schematics of ion accumulation under negative and positive bias.
\label{cation-anion}}
\end{figure*}

The temperature-dependent current transients for a negative gate pulse (0 to --2.5 V, 20 s) shown in \fig{cation-anion}(a) reveal a striking progression in ionic mobility (the same for positive gate pulse 0 to 3 V is depicted in Fig. S3 \cite{supple}). At high temperatures ($\geq$260 K), the normalized drain current rises rapidly following a gate-voltage pulse, indicating that ions reorganize quickly enough to track the applied field. The corresponding relaxation is rapid and broadly distributed, consistent with a liquid that supports many rearrangement pathways. As temperature decreases, the transient becomes stretched and slow, exhibiting incomplete equilibration even over extended timescales. A small number of relaxation modes remain active as the liquid approaches a dynamically arrested state. To capture both ionic species (EMIM$^+$ and TFSI$^-$), \fig{cation-anion}(b) employs an alternating-polarity gating scheme consisting of periodic negative (0 $\rightarrow$ –2.5 V) and positive (0 $\rightarrow$ +3 V) pulses. The negative step drives EMIM$^+$ toward the interface, whereas the positive step drives the bulkier TFSI$^-$ anion. The resulting drain-current waveform clearly distinguishes the two relaxation processes: EMIM$^+$ responds quickly, while TFSI$^-$ exhibits a substantially longer relaxation time. This asymmetry reflects the larger hydrodynamic radius and reduced mobility of TFSI$^-$, demonstrating that the transistor directly resolves ion-specific kinetics. Relaxation times $\tau(T)$ for both ions are extracted by fitting each transient to the Kohlrausch-Williams-Watts (KWW) stretched-exponential form \cite{Vela2020JCP},
\begin{equation}
I_{\text{ds}}(t) = I_0\bigg[1-\exp\bigg(-\Big(\frac{t}{\tau}\Big)^\beta\bigg)\bigg]
\label{KWW}
\end{equation}
as shown in \fig{cation-anion}(c). Across all temperatures, $\tau_{\text{EMIM}^+}$ $<$ $\tau_{\text{TFSI}^-}$, consistent with the relative mobilities inferred from \fig{cation-anion}(b). The inset displays a representative ambipolar transfer curve and gating schematics that illustrate how the polarity of the applied gate bias determines whether EMIM$^+$ or TFSI$^-$ approaches the PdSe$_2$ surface. As seen from \fig{cation-anion}(c), extracted $\ln \tau$ for both EMIM$^+$ and TFSI$^-$ deviates from conventional Arrhenius behavior, but rather follows the Vogel–Fulcher-Tammann (VFT) scaling \cite{Angell2000JAP,Tinoco2016SciRep},
\begin{equation}
\tau(T) = \tau_0 \ \exp\bigg(\frac{B}{T-T_0}\bigg),
\label{VFT}
\end{equation}
which is a classical signature of glass-forming systems and describes their super-Arrhenius relaxation dynamics with a divergence at a finite temperature. This can also be written in a modified form \cite{Angell1995Science},
\begin{equation}
\tau(T) = \tau_0 \ \exp\bigg(\frac{DT_0}{T-T_0}\bigg),
\label{modified-VFT}
\end{equation}
where $\tau_0$, $B$, and $T_0$ are material-dependent constants. $T_0$ defines the extrapolated diverging point of relaxation time, often called Kauzmann temperature, which is thought to be the ideal glass temperature, usually 20–50 K below $T_\text{g}$. The parameter $D$ controls how closely the system obeys the Arrhenius law ($D = \infty$). Another parameter called fragility index ($m$) can be extracted from the slope of ``Angell plot", $\log_{10}\tau$ vs $T_g/T$, and interprets how quickly the dynamics of glass-forming liquid slows down upon cooling \cite{Angell1995Science,Jaiswal2016PRL}. ``Strong" liquids like SiO$_2$ show an Arrhenius variation of structural relaxation time between the glass transition temperature $T_\text{g}$ and high temperature limits. For the other extreme, ``fragile" liquids, having glassy state structures reorganized with little provocation from thermal excitation, the relaxation time varies in a strongly non-Arrhenius fashion between high and low limits. The strong or fragile liquids can be characterized by their strength parameter $D$, which typically attains values less than 10 for fragile liquids \cite{Angell1995Science}. 
Values of $D$ and $T_0$ extracted for EMIM$^+$ are 3.14 and 156 K, respectively and that for TFSI$^-$ are 2.54 and 160 K, respectively. The faster EMIM$^+$ cation, having access to more microstates and lower steric hindrance, reaches its VFT divergence at lower temperature. This ion-specific glassy dynamics highlights the sensitivity of the method to microscopic asymmetries in ionic motion. The fragility index ($m$) is defined as \cite{Jaiswal2016PRL,Böhmer1993JCP},
\begin{equation}
m = \left. \frac{d\,\log_{10} \ \!\tau(T)}{d\,(T_\text{g}/T)} \right|_{T = T_\text{g}}
\label{frag-slope}
\end{equation}
Combining \eqref{modified-VFT} and \eqref{frag-slope},
\begin{equation}
m = \frac{DT_0T_\text{g}}{\ln 10 \ (T_\text{g}-T_0)^2}
\label{frag-D}
\end{equation} 
From the VFT fitting of $\tau_{\text{EMIM}^+}$ and $\tau_{\text{TFSI}^-}$, the estimated $m$ values are 98 and 142, respectively, considering the calorimetric glass transition temperature ($T_\text{g}$) for EMIM-TFSI as 175 K \cite{Zhang2006JPCrefdata}. The fits place the ionic liquid firmly within the fragile regime of Angell’s classification (for strong liquids like SiO$_2$, $m$ $<$ 20) \cite{Angell1995Science}.
 
\begin{figure}[t]
\centerline{\includegraphics[width=\textwidth, clip]{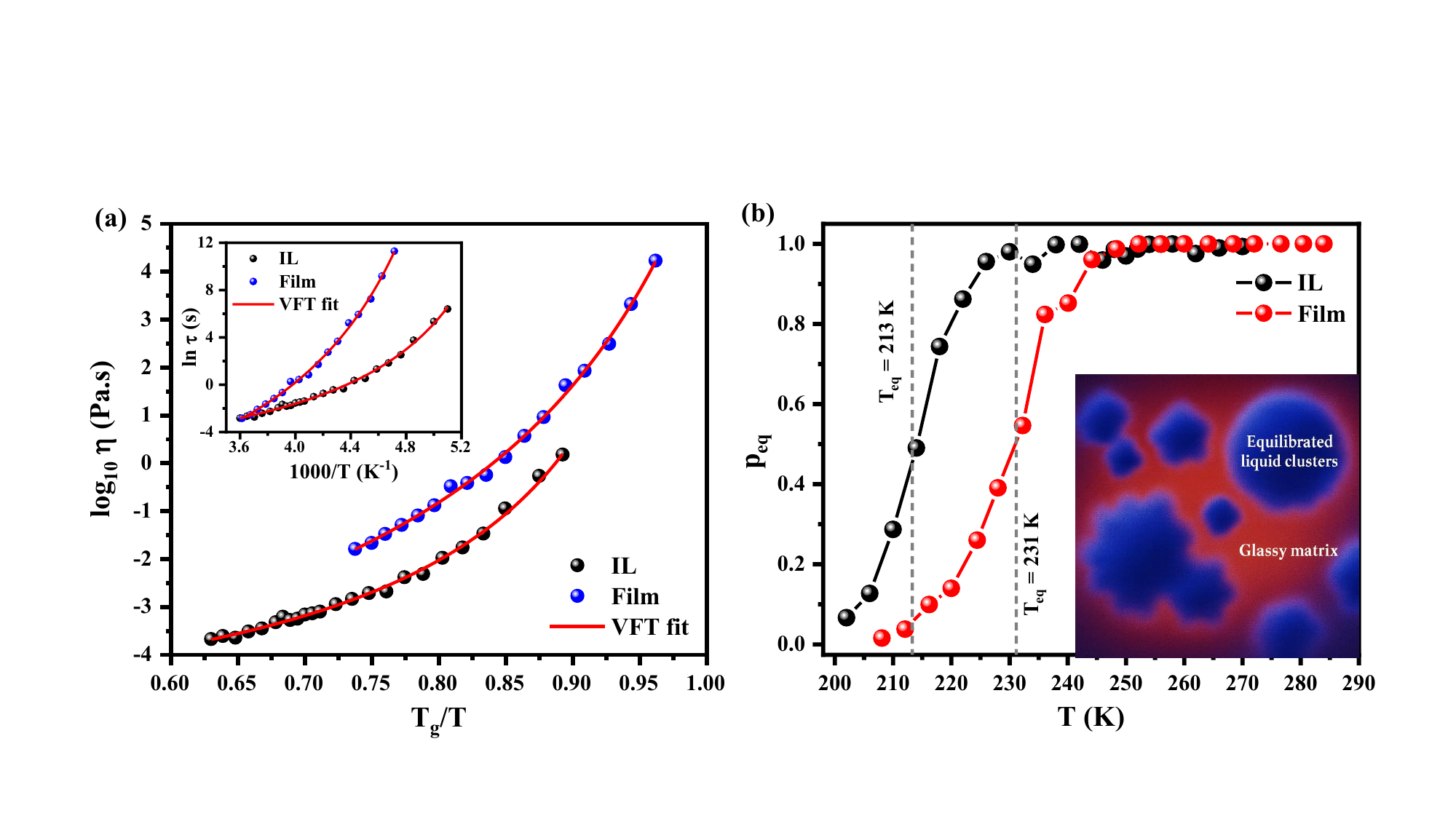}}
\caption{(a) Angell plot of viscosity ($\log_{10}\eta$ vs. $T_\text{g}/T$) for both the ionic liquid and its PVDF–HFP polymerized ionogel. Inset shows the corresponding $\ln\tau$ vs 1000/$T$ plot. (b) Temperature dependence of the equilibrated fraction $p_{\text{eq}}$ for the liquid and ionogel, revealing a sharp collapse towards glass transition. Dashed lines mark the characteristic temperatures where 50\% equilibrated domains coexist with the glassy phase. Inset shows the schematic of equilibrated clusters embedded in a glassy matrix. 
\label{comparison}}
\end{figure}

The most significant rheological parameter, viscosity ($\eta$), can be semiquantitatively inferred in a simplified approach from the measured ionic relaxation times by combining the classical Stokes-Einstein (SE) relation with a diffusion estimate derived from the electrical transients,
\begin{equation}
\eta(T) = \frac{k_BT}{6 \pi rD^\prime(T)} \
\label{viscosity}
\end{equation}
where $k_B$ is Boltzmann’s constant, $r$ is an effective hydrodynamic radius for the mobile ion ($\sim$ 0.33--0.38 nm) \cite{Qu2022ACSomega}, and $D^\prime$ is the diffusion constant. $D^\prime$ can be estimated using Fick's diffusion law, $D^\prime(T) = \frac{L^2}{6\tau}$, where $\tau(T)$ is the KWW-extracted relaxation time, and $L$ is the diffusion path length. The factor 6 arises from the linear growth of the mean square displacement $\langle r^2(t) \rangle$ = 6$D^\prime t$ for isotropic three-dimensional diffusion (although the electronic transport is confined to a two-dimensional channel, the ionic motion responsible for screening the gate field involves transport through the bulk of the ionic liquid or ionogel and is therefore treated as three-dimensional). The Stokes–Einstein relation is used here as a simplified approach to extract viscosity, although it assumes hydrodynamic, continuum-like diffusion of roughly spherical particles, whereas deeply supercooled ionic media often show fractional Stokes–Einstein behavior and strong dynamic heterogeneity \cite{Hodgdon1993PRE,Pan2017SciRep}. We apply this procedure to both EMIM–TFSI liquid and its PVDF–HFP polymerized ionogel film in which the ionic liquid is integrated within a semi-crystalline polymer scaffold. Ionogel-gated transfer curves at different temperatures are depicted in Fig. S4 \cite{supple}. The ionogel imposes reduced free volume and specific ion–polymer coordination, both of which increase the energetic barriers for ion hops and enlarge the cooperative volume required for relaxation. Consistently, the ionic relaxation becomes slower and the extracted $\tau$ as well as $\eta(T)$ for the film are systematically larger and display steeper VFT slope as shown in the $\log_{10}\eta$ vs. $T_\text{g}/T$ plot in \fig{comparison}(a) (only the cationic contribution for liquid and film are represented here for comparison). 

However, in this work, Fick's law estimate $D^\prime(T) = \frac{L^2}{6\tau}$ is used to define an effective diffusion length over which the collective ionic polarization relaxes in our device geometry, not the microscopic mean-free-path of individual ions. For the ionic liquid droplet configuration, $L \sim$30 $\mu$m is the gate-to-channel separation that governs the long-wavelength relaxation mode of the EDL charging process, as described within the Poisson-Nernst-Planck framework, whereas in the case of ionogel film, $L$ is the thickness of the film, $\sim$3.4 $\mu$m. In both cases, the extracted $D^\prime(T)$ carries the same temperature dependence as $\tau(T)$ and provides a consistent relative comparison between the two systems. Therefore, the fractional SE contribution would rescale the absolute magnitude of the diffusion constant, but would not affect the observed temperature dependence, VFT behavior, or comparative trends between liquid and ionogel systems. 

Within this simplified framework, the electrically inferred viscosity of the ionogel exceeds that of the neat ionic liquid by nearly two orders of magnitude at 280 K and increases further upon cooling as the two systems approach their respective glass transitions at different rates. The absolute values are likely underestimated using the classical SE relation rather than the fractional SE formalism expected for glass-forming systems. The inset of \fig{comparison}(a) shows that $\ln\tau$ vs. 1000/$T$ for both systems follows VFT behavior, corroborating that polymer coupling accelerates ergodicity loss in the ionogel. The representative drain current transients for a negative gate pulse in the film at different temperatures are displayed in Fig. S5(a) \cite{supple}.
From the VFT fitting of the Angell's plot of ionogel film, the strength parameter $D$, diverging temperature $T_0$, and fragility index $m$ are estimated to be 5.63, 170 K, and 73, respectively, considering the value of calorimetric glass transition temperature (T$_\text{g}$) for the ionogel as 204 K \cite{Qu2022ACSomega}. While the composite system exhibits a higher $T_\text{g}$ than the pure ionic liquid, its lower fragility index indicates a transition toward a ``stronger" glass-forming behavior. A higher VFT divergence temperature $T_0$ is detected for the film relative to the neat liquid, reflecting the reduced configurational entropy accessible to ions inside the polymer scaffold. According to Angell's classification, the ionogel film serves as a moderate fragile glass former.

Since glasses are out-of-equilibrium systems, they will evolve towards equilibrium by annealing to a certain temperature. Different regions in the glass will follow a distribution of relaxation times depending on their mobility and external temperature and thereby produce a stretching relaxation signature. This dynamically arrested state near the glass transition can be described as the combination of spatial pockets that equilibrate within the experimental timescale coexisting with the surrounding glassy state [inset of \fig{comparison}(b)] \cite{Costa2022NatPhy}. When a glass is heated to temperatures far above its glass temperature, the emergence of such equilibrated liquid regions is much faster than the relaxation of the glass and facilitates the transition of the glass into liquid. To quantify the equilibrated liquid regions at a certain temperature we introduce an electrically accessible dynamical quantity $p_{\text{eq}}(T)$, representing the fraction of the ionic medium that equilibrates within the experimental timescale. This quantity is computed directly from the KWW-extracted parameters $\tau(T)$ and $\beta$ and provides an operational measure of ergodic–nonergodic crossover that is accessible through transistor drain-current transients,
\begin{equation}
p_{\text{eq}}(T) = 1 - \exp\bigg[-\bigg(\frac{t_{\text{obs}}}{\tau(T)}\bigg)^\beta\ \bigg]
\label{equilibrated}
\end{equation}  
where $t_{\text{obs}}$ is the characteristic measurement window (here 20 s). For $\tau(T)$ $<<$ $t_{\text{obs}}$ (at high temperature), the system equilibrates fully and $p_{\text{eq}}$ $\approx$ 1. As the system approaches its glass crossover, $\tau$ grows larger than the observation window, and only a fraction of microscopic regions can relax, therefore corresponding $p_{\text{eq}}$ collapses sharply (\fig{comparison}(b)). The ionogel exhibits an earlier and more pronounced decline in $p_{\text{eq}}$, indicating that the polymer matrix restricts the connectivity of dynamically active pathways. We define a characteristic temperature ($T_\text{eq}$) denoting the approximate onset of non-ergodic dynamics, where $p_\text{eq}$ attains a value 0.5. The experimental estimate of $T_\text{eq}$ is $\sim$ 213 K for neat EMIM-TFSI and $\sim$ 231 K for its polymerized ionogel.

The physical picture of $p_\text{eq}$ can be well analyzed from percolation theory. At high temperature, equilibrating domains percolate throughout the system, forming a connected network of mobile clusters. As temperature decreases, the network fragments into isolated fractal islands embedded in a rigid, glassy matrix \cite{Wool2008JPS}. When $p_\text{eq}$ approaches the percolation threshold, long-range connectivity is lost and the system can no longer explore its configurational phase space within any reasonable experimental timescale, consistent with dynamical facilitation theory \cite{Chacko2024PRX}, in which mobility propagates through correlated excitations that become increasingly sparse upon cooling. In the ionogel, polymer hinders the excitation propagation, shifting the percolation threshold to higher temperature. According to Wool's formalism, the onset of long-range connected fractal structures begins at a characteristic temperature $T^*$, that lies above the calorimetric glass transition, typically $T^*$ $\approx$ 1.2$T_\text{g}$ \cite{Wool2008JPS,Stanzione2011J.NonCrysSolids}. In \fig{comparison}(b), the characteristic temperature $T_\text{eq}$ may be viewed as a dynamical analog of $T^*$, with $T_\text{eq}$/$T_\text{g}$ = 1.21 for ionic liquid and 1.13 for ionogel, marking the onset of fractal dynamics well above the glass transition.

Within the Adam–Gibbs (AG) framework \cite{Adam1965JCP}, the rapid collapse of the fraction of equilibrated liquid regions arises from the progressive loss of configurational entropy $S_c(T)$, which quantifies how many ways the system can rearrange itself, i.e., the degree of freedom of structural relaxation. It can be linked with the relaxation dynamics as,
\begin{equation}
\tau(T) = \tau_0\exp\bigg(\frac{C}{TS_c(T)}\bigg)
\label{conf.entropy}
\end{equation}
Considering VFT law mathematically equivalent to AG's law, $S_c(T)$ can be scaled as, $S_c(T) \propto \frac{T-T_0}{T}$. At high temperature, ions can freely explore many configurations and provides a large $S_c$. Near the glass transition the ions get trapped in a dynamically arrested state and $S_c$ $\rightarrow$ 0. The diverging temperature $T_0$, therefore, represents the zero configurational entropy limit. 
As in the present device ionic liquid is confined within the EDL, the accessible configurational phase space is reduced relative to the bulk \cite{Richert2011AnnRev}. This entropy suppression manifests an earlier loss of ergodicity, shifting the glass transition and the collapse of $p_{eq}$ to higher temperatures. A phenomenological scaling for the confined entropy is described in the Supplemental Material.

\section{Conclusions}
\label{sec:conclusion} 

In this work, we establish an ionic-gated van der Waals transistor as a semi-quantitative electrical probe of supercooled ionic dynamics beyond the reach of conventional rheometry. Ambipolar PdSe$_2$ FET resolves ion-specific relaxation, viscosity growth, and the onset of nonergodicity through purely electrical measurements. The resulting signals reveal hallmark features of fragile glass formation including stretched-exponential relaxation, Vogel–Fulcher–Tammann scaling, and the fragmentation of mobile domains upon cooling. Polymer confinement shifts the characteristic glassy temperatures upward, highlighting the sensitivity of the method to structural constraints imposed by the polymer matrix. This approach enables rheological characterization at the picoliter scale within functional solid-state architectures, providing a direct route to probe nonequilibrium glass physics in device-integrated environments.

\begin{acknowledgments}
T.K. acknowledges DST-INSPIRE and IACS for fellowship support. R.P. acknowledges CSIR, S.M. acknowledges UGC, A.S., A.S. and B.K. acknowledge IACS for fellowship support. T.K. and S.D. would like to thank Prof. Biman Jana for useful discussions. T.K. acknowledges Dr. Urmimala Maitra for experimental help. S.D. acknowledges the financial support from DST-ANRF under grant No. CRG/2021/004334. S.D. also acknowledges support from the Technical Research Centre (TRC), IACS, Kolkata.

\end{acknowledgments}

\section*{Data Availability}

The data that supports the findings of this article cannot be made publicly available. The data are available upon reasonable request from the authors.

\newpage

\end {document}


\title{Supplementary Information:\\
Electronic access to glass transition in supercooled ionic liquids using ambipolar transistor}

\author{Tanima Kundu}
\affiliation{School of Physical Sciences, Indian Association for the Cultivation of Science, 2A $\&$ B
Raja S. C. Mullick Road, Jadavpur, Kolkata - 700032, India
}

\author{Rahul Paramanik}
\affiliation{School of Physical Sciences, Indian Association for the Cultivation of Science, 2A $\&$ B
Raja S. C. Mullick Road, Jadavpur, Kolkata - 700032, India
}

\author{Aishee Saha}
\affiliation{School of Physical Sciences, Indian Association for the Cultivation of Science, 2A $\&$ B
Raja S. C. Mullick Road, Jadavpur, Kolkata - 700032, India
}

\author{Shibnath Mandal}
\affiliation{School of Physical Sciences, Indian Association for the Cultivation of Science, 2A $\&$ B
Raja S. C. Mullick Road, Jadavpur, Kolkata - 700032, India
}

\author{Antarik Sarkar}
\affiliation{School of Physical Sciences, Indian Association for the Cultivation of Science, 2A $\&$ B
Raja S. C. Mullick Road, Jadavpur, Kolkata - 700032, India
}

\author{Bipul Karmakar}
\affiliation{School of Physical Sciences, Indian Association for the Cultivation of Science, 2A $\&$ B
Raja S. C. Mullick Road, Jadavpur, Kolkata - 700032, India
}

\author{Subhadeep Datta}
\email{sspsdd@iacs.res.in}
\affiliation{School of Physical Sciences, Indian Association for the Cultivation of Science, 2A $\&$ B
Raja S. C. Mullick Road, Jadavpur, Kolkata - 700032, India
}

\maketitle

\section*{Device architecture and gating response}

Multilayered PdSe$_2$ flakes were contacted with Ti/Au electrodes and the ionic liquid (EMIM-TFSI) was dropped onto the channel covering the source, drain, and gate electrodes (manuscript Fig. 1(a)). Applying an ionic-gate voltage drives EMIM$^+$ or TFSI$^-$ ions toward the semiconductor surface, where they assemble into a nanometer-scale electric double layer (EDL). This interfacial capacitor, with several $\mu$F cm$^{-2}$ capacitance ($C_{\text{EDL}}$), enables carrier modulation far beyond that achievable with conventional oxide dielectrics ($e.g.$ SiO$_2$) \cite{Fujimoto2013pccp}. An equivalent circuit model shown in manuscript Fig. 1(a) depicts the combined action of the EDL capacitance ($C_{\text{EDL}}$) and the back-gate oxide capacitance ($C_{\text{ox}}$). The resulting electrostatic tunability is manifested in the color map in manuscript Fig. 1(b) generated by the back-gated transfer curves ($I_{\text{ds}}$ vs. $V_{\text{bg}}$) measured under different ionic-gate biases ($V_{\text{ig}}$). The corresponding stack plot shown in Fig. S1 clearly portrays the evolution of back-gated ambipolar transport towards hole-dominated (\emph{p} type) transport under negative ionic-gate bias and electron-dominated (\emph{n} type) transport under positive ionic-gate bias at room temperature, which is basically caused by the accumulation of positive (negative) ions at the channel interface when $V_{\text{ig}}$ $<$ 0 ($V_{\text{ig}}$ $>$ 0). The back-gated charge neutrality point (CNP) $V_\text{CNP}$ (extracted from the forward sweep) is plotted against $V_{\text{ig}}$ at room temperature and near glass transition (200 K) (manuscript Fig. 1(c)). Inset shows an representative ambipolar transfer curve of PdSe$_2$ indicating the charge neutrality points in both forward and backward sweeps. While $V_\text{CNP}$ is strongly tunable by $V_{\text{ig}}$ at room temperature, this tunability is substantially suppressed upon cooling toward the glass transition. In this regime, $V_\text{CNP}$ becomes nearly insensitive to $V_{\text{ig}}$, consistent with the partial freezing of ionic motion. Consequently, increasingly large gate voltages are required to modulate the CNP as the ionic liquid approaches a glassy state.

\subsection*{Static gating sweeps}

Ionic-gated transfer curves ($I_{\text{ds}}$ vs. $V_{\text{ig}}$) are measured at different temperatures (Fig. S2). The corresponding hysteresis width $\Delta V_\text{H}$ serves as an indicator of fast or slow response of mobile ions. At high temperature, where the ionic liquid remains fully ergodic, the hysteresis loop is narrow and ions fully relax within each sweep, as shown in the 300 K transfer curve. The direction of hysteresis anti-clockwise on the \emph{n} side and clockwise on the \emph{p} side reflects the finite response time of ionic rearrangement within the EDL during gate reversal. On cooling, as the ionic relaxation becomes slower than the gate bias sweep rate, the system falls temporarily out of equilibrium, producing a lagging polarization that appears as a loop in the transfer curve. Near the glass-transition regime, however, the hysteresis expands markedly, reflecting slow or partially frozen ionic motion, as shown in the 200 K trace. The transition in the transfer curve morphology from a sharp, reversible gating response at high temperature to a broadened and distorted response at low temperature signals the contraction of the dynamically active mobile networks within the ionic liquid. The asymmetric magnitude of hysteresis between the electron- and hole-conduction branches points to unequal ionic mobility or interfacial charge-trapping dynamics at opposite polarities. 

\subsection*{Time-resolved dynamics}

Drain-current transients ($I_{\text{ds}}(t)$) were recorded by applying sudden voltage steps on the ionic gate which is predominantly contributed by the ionic relaxation \cite{Li2025AdvMatTech,Zhao2024NanoMicroLett}. In a minimal electrical description, the characteristic timescale associated with transport-limited charging of the EDL is given by the $RC$ time constant of ionic gate circuit, $\tau_\text{e} = R_{\text{IL}}\ C_{\text{EDL}}$, where $R_{\text{IL}}$ represents the resistive contribution arising from bulk ion transport. Beyond this electrical picture, the viscoelastic nature of the ionic liquid can be described by a rheological equivalent circuit based on the Maxwell's spring-dashpot model. Within this framework, ionic relaxation is governed by both viscous damping, characterized by the viscosity ($\eta$), and elastic response, quantified by the shear modulus $\zeta$. This viscoelastic contribution introduces an additional relaxation timescale, $\tau_\text{r}$ = $\frac{\eta}{\zeta}$, associated with viscosity, a measure of the liquid response to a suddenly imposed shear stress. Consequently, the observed current transient does not reflect a single universal time constant, but rather the interplay between transport-limited EDL charging and viscoelastic relaxation in the ionic liquid. The transients are fitted using a stretched exponential. A finite relaxation time $\tau$ signifies measurable ionic rearrangement, enabling extraction of $\tau$ and the stretch exponent $\beta$, followed by continued measurements upon cooling. When no finite exponential relaxation is detectable, $\tau$ diverges, signaling glassy arrest. 

Together, these static and dynamic measurements enable a self-referenced extraction of ionic dynamics leading to interpretation of viscosity, fragility, and the emergence of percolating fractal dynamics. 

\section*{Configurational entropy}

In confined or interfacial environments, theoretical studies have shown that configurational entropy is further suppressed due to finite-size effects and restricted cooperativity \cite{Richert2011AnnRev}. In the present device-integrated geometry, the ionic liquid is effectively confined within a nanoscale electric-double-layer region, and the accessible configurational phase space is therefore reduced relative to the bulk. As the liquid is supercooled, the structural correlation length, $\xi(T)$, representing the size of cooperatively rearranging regions (CRRs), grows significantly \cite{Kirkpatrick1989PRA}. Although $S_c$ and $\xi$ are not directly measurable in our transport experiments, we can adopt a phenomenological scaling for the confined entropy \cite{Keddie1994EPL,Alcoutlabi2005JPCM},
\begin{equation}
S_c^\text{confined}(T) \approx S_c^\text{bulk}(T) \Big[1-\Big(\frac{\xi(T)}{\lambda}\Big)^\gamma\Big]
\end{equation}
where $\gamma$ is a geometric exponent, and $\lambda$ is an effective confinement thickness. As $\xi(T)$ approaches the confinement length $\lambda$, the configurational entropy is truncated, leading to a precipitous drop in the number of available microstates compared to the bulk. This entropy suppression manifests an earlier loss of ergodicity, shifting the glass transition and the collapse of the dynamical order parameter, $p_{eq}(T)$, to higher temperatures. Consequently, the transistor-based rheometer captures the thermodynamic frustration of ions at the nanoscale, where the geometric restriction of the EDL acts as a catalyst for glassy arrest.

\newpage

\begin{figure}[ht!]
\centerline{\includegraphics[scale=0.8, clip]{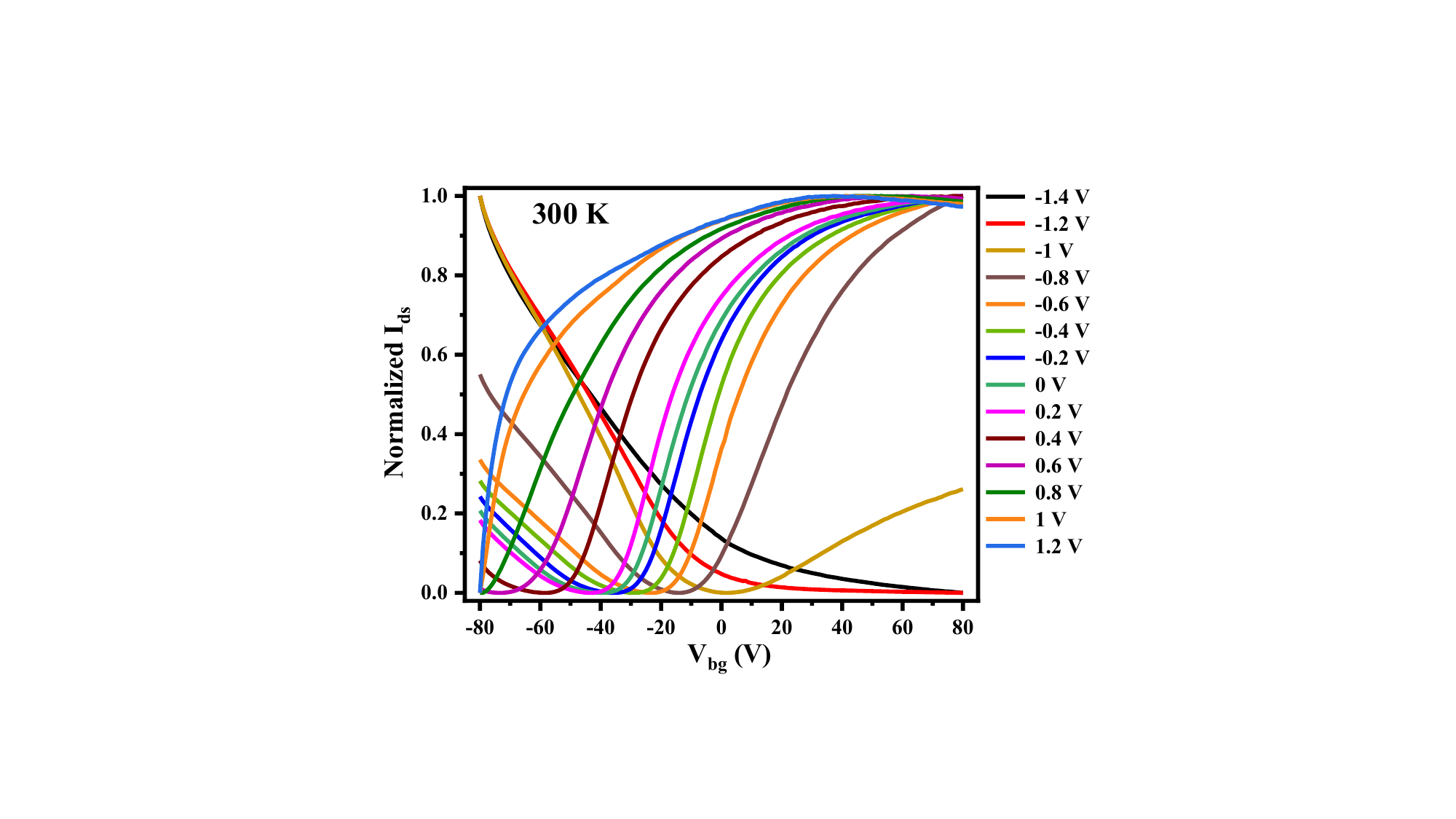}}
\renewcommand{\thefigure}{S\arabic{figure}}
\caption{Back-gated transfer curves ($I_\text{ds}$ vs. $V_\text{bg}$) of PdSe$_2$ at different ionic gate bias ($V_\text{ig}$) considering only forward sweep (-80 V to +80 V) at room temperature.
\label{stack plot}}
\end{figure}

\begin{figure}[ht!]
\centerline{\includegraphics[scale=0.7, clip]{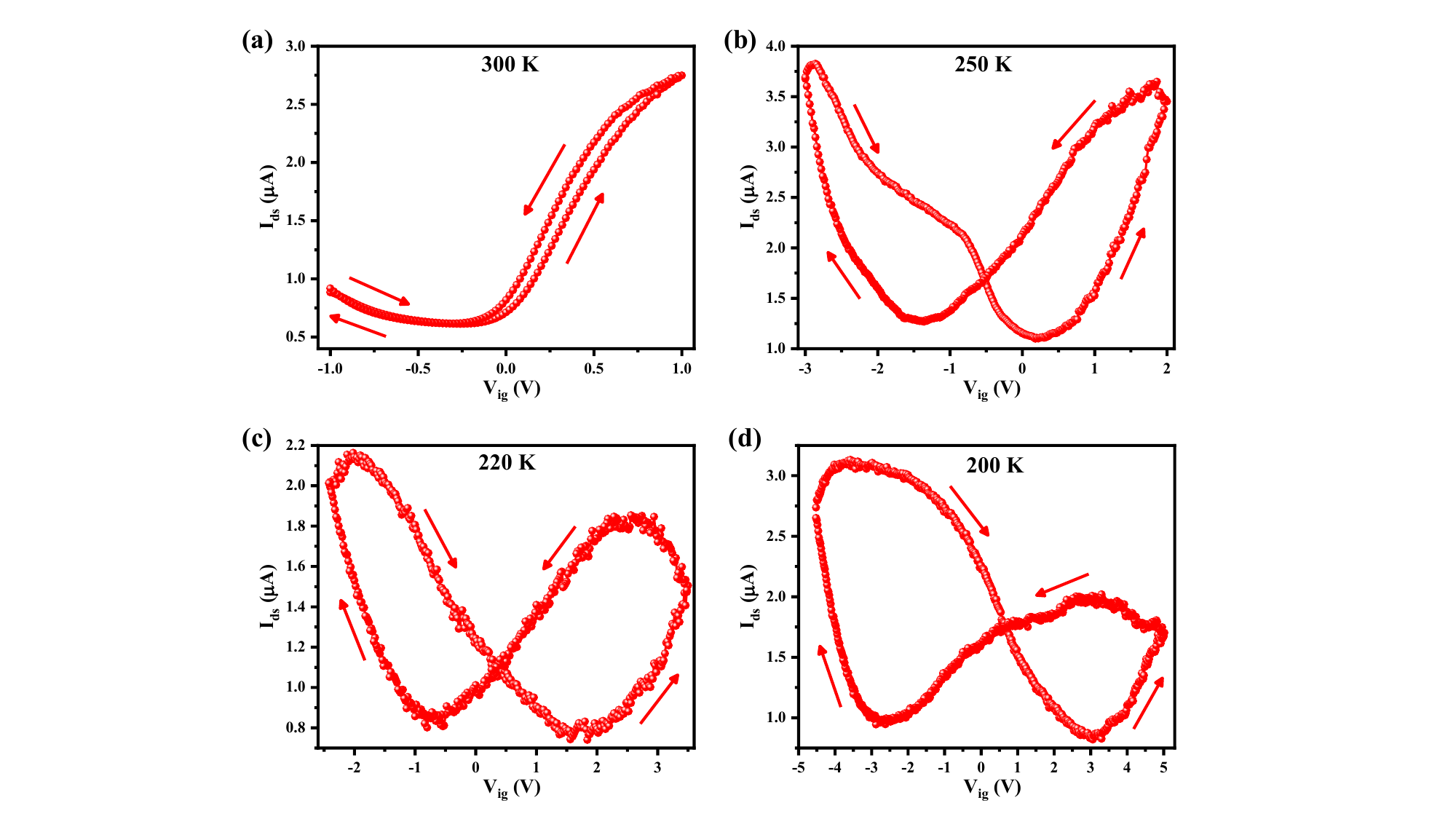}}
\renewcommand{\thefigure}{S\arabic{figure}}
\caption{(a)-(d) Ionic-liquid-gated transfer curves ($I_\text{ds}$ vs. $V_\text{ig}$) of PdSe$_2$ FET at different temperature, showing a transition from narrow hysteresis and reversible gating response at high temperature to a broadened hysteresis on approaching towards glass transition (around 200 K), reflecting lagging polarization and gradual freezing of ionic motion. The arrows indicate the direction of hysteresis.
\label{IL gating}}
\end{figure}

\begin{figure}[ht!]
\centerline{\includegraphics[scale=0.8, clip]{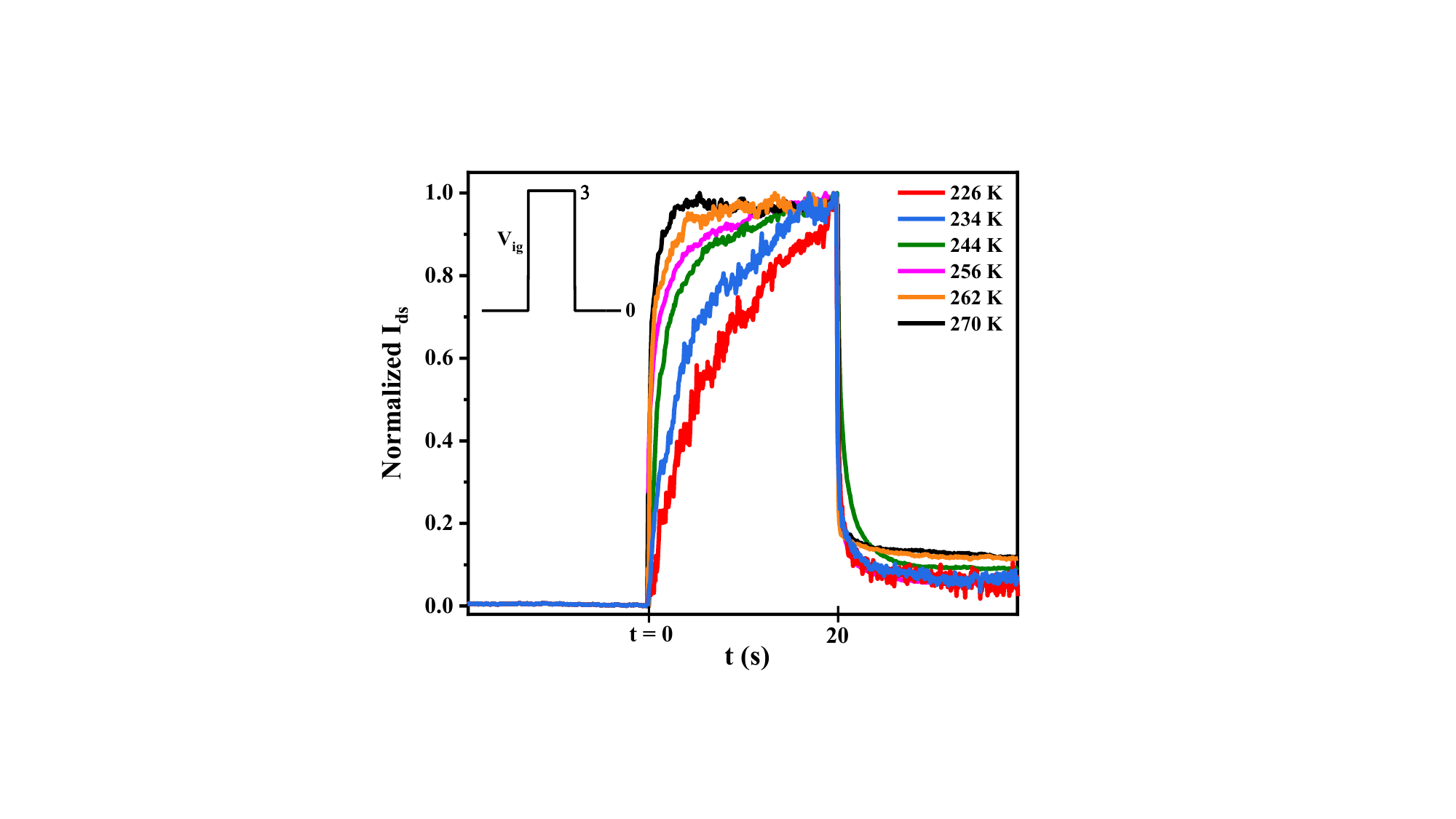}}
\renewcommand{\thefigure}{S\arabic{figure}}
\caption{Temperature-dependent drain current transients recorded by applying a positive pulse from 0 to 3 V on the ionic-liquid gate, showing a progressive slowdown of the ionic (TFSI$^-$) response upon cooling.
\label{positive pulse}}
\end{figure}

\begin{figure}[ht!]
\centerline{\includegraphics[scale=0.6, clip]{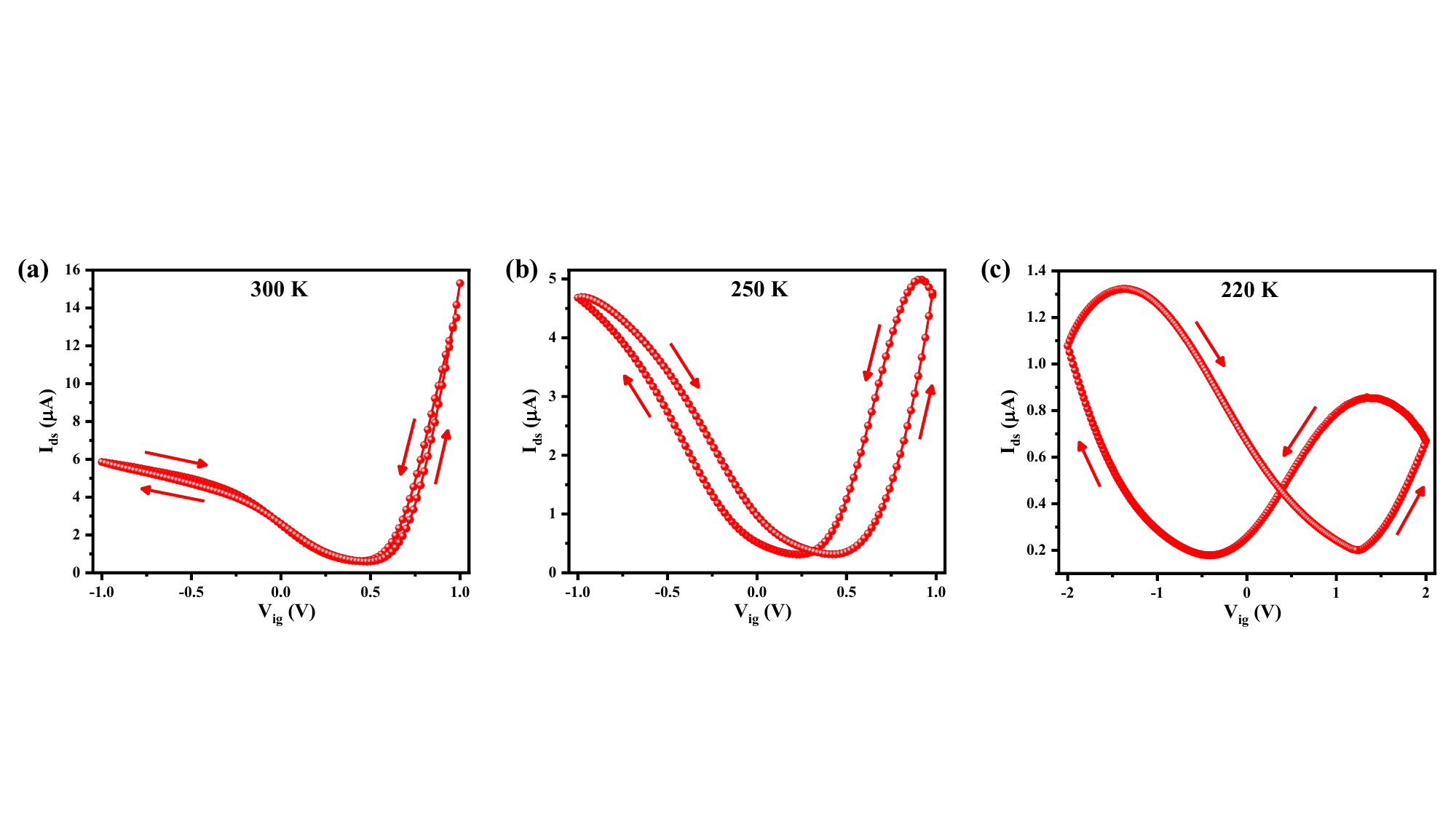}}
\renewcommand{\thefigure}{S\arabic{figure}}
\caption{(a)-(c) Ionogel-gated transfer curves ($I_\text{ds}$ vs. $V_\text{ig}$) of PdSe$_2$ FET at different temperature, showing a transition from narrow hysteresis and reversible gating response at high temperature to a broadened hysteresis on approaching towards glass transition (around 220 K), reflecting lagging polarization and gradual freezing of ionic motion within polymer confinement. The arrows indicate the direction of hysteresis.
\label{Film gating}}
\end{figure}

\begin{figure}[ht!]
\centerline{\includegraphics[scale=0.6, clip]{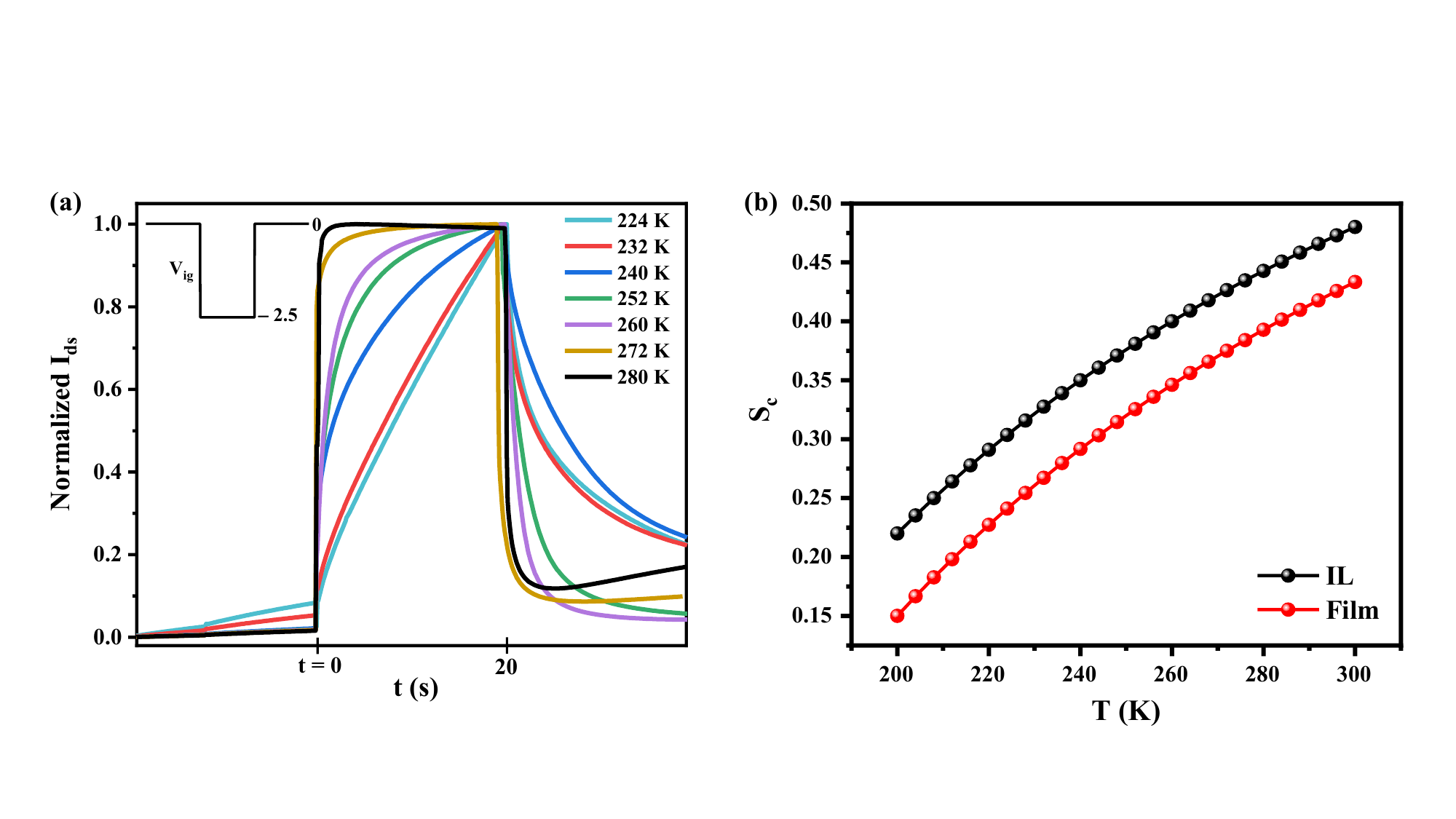}}
\renewcommand{\thefigure}{S\arabic{figure}}
\caption{(a) Temperature-dependent drain current transients recorded by applying a negative pulse from 0 to -2.5 V on the ionogel top gate, showing a progressive slowdown of the ionic response upon cooling. (b) Scaling of configurational entropy ($S_c$ $\propto$ $\frac{T-T_0}{T}$) with temperature for both ionic liquid and ionogel film, revealing the gradual decrease of $S_c$ on cooling. Near the glass transition $S_c$ $\to 0$ due to the freezing of ionic motion in a dynamically arrested glassy matrix.   
\label{film transient and conf. entropy}}
\end{figure}